\documentclass[12pt]{article}

\begin{document}

\title{Quantization via hopping amplitudes: \\
       Schr\"odinger equation and free QED\\[10mm]}

\author{{\bf L. Polley}\\[5mm]
FB Physik\\Oldenburg University\\26111 Oldenburg, Germany\\[2mm]
e-mail: {\tt polley@uni-oldenburg.de}\\[10mm]}
\maketitle\vspace{10mm}
\begin{abstract}
Schr\"odinger's equation with scalar and vector potentials is shown to describe 
``nothing but'' hopping of a quantum particle on a lattice; any spatial variation
of the hopping amplitudes acts like an external electric and/or magnetic field. 
The main point of the argument is the superposition principle for state vectors; 
Lagrangians, path integrals, or classical Hamiltonians are not (!) required. 
Analogously, the Hamiltonian of the free electromagnetic field is obtained as 
a twofold continuum limit of unitary hopping in $Z(N)$ link configuration space, 
if gauge invariance and $\cal C$ and $\cal P$ symmetries are imposed. 
\\[2mm]
\end{abstract}
\thispagestyle{empty}

\newpage\section{Introduction}

For at least two standard quantum systems, 
canonical quantization (or other classical-to-quantum substitution rules) 
can be avoided; it can be replaced by an intrinsic quantum mechanical 
consideration of ``hopping'' in a discrete configuration space. 
This only requires to interpret a familiar tool from model 
building---hopping amplitudes---as a first-principle concept.  
   
Hopping amplitudes have a long tradition particularly in 
solid-state theory \cite{Heisenb,Hubbard}. On a fundamental level they 
have been used in lattice gauge theory \cite{Kogut,Creutz} for discretizing 
(not avoiding) path-integral actions. 
More recently, in the field of quantum computation, hopping parameters are being 
used as collision constants in unitary cellular automata \cite{Bialyn,Boghosian} 
designed for efficient simulation of the Schr\"odinger equation
\cite{Boghosian} or 1-photon and Weyl equation \cite{Bialyn}. These latter
applications differ in a crucial way from the viewpoint taken here, by 
assuming locality in conjunction with a finite, irreducible time step. 
It has proven to be a major challenge to design algorithms satisfying that 
computational requirement. Apart from technical complications, however, 
unitary cellular automata in some cases require configuration spaces 
larger than the physical ones. For example, local hopping rules in 
$d$ spatial dimensions are found to require $2d$-component wave functions 
\cite{Boghosian}. Consequently, a {\em real} spinless particle (as opposed to 
its computer simulation) can have a unitary and local equation of motion only 
with respect to {\em continuous time}. 

Hopping amplitudes can do more than approximate or discretize 
processes originally defined otherwise. They necessarily emerge as coefficients 
of a superposition when a particle is prepared in a position eigenstate. 
The crucial axiom here is that the state of a quantum particle is
completely specified by a {\em position at one instant of time}. To illustrate
the idea, consider a particle confined to a 1-dimensional array of discrete 
positions at a spacing $a$. Let us work in the Heisenberg picture and denote by 
$|n,t\rangle$ the eigenstate of position $x=na$ at time $t$. 

To prepare a position $n$ at time $t$ means to prepare a state with an 
uncertain position at time $t+{\rm d}t$, because any motional information is 
lacking from $|n,t\rangle$. For ${\rm d}t$ small enough, the
uncertainty only relates to positions $n$, $n+1$, and $n-1$. 
Furthermore, $n+1$ and $n-1$ will occur symmetrically if we assume the 
symmetries of a free particle. Thus 
\begin{equation}     \label{nHop} 
   |n,t\rangle = \alpha |n,t+{\rm d}t\rangle 
               + \beta |n+1,t+{\rm d}t\rangle + \beta |n-1,t+{\rm d}t\rangle 
\end{equation}
where $\alpha$ and $\beta$ are some numbers dependent on the 
size of the time step. For ${\rm d}t\to 0$ we must have 
$\alpha\to 1$ and $\beta\to 0$, hence
$$ 
    \alpha = 1 + \alpha_1 {\rm d}t + {\cal O}({\rm d}t^2)  \qquad \qquad
    \beta  =     \beta_1 {\rm d}t + {\cal O}({\rm d}t^2)
$$
Thus the basic hopping equation (\ref{nHop}) converges to the differential
equation 
\begin{equation}     \label{diffHop}
                -\frac{\rm d}{{\rm d}t} |n,t\rangle = \alpha_1 |n,t\rangle
      + \beta_1 |n+1,t\rangle + \beta_1 |n-1,t\rangle 
\end{equation}
We now use the statistical interpretation of the scalar product. From 
$$
   \langle n, t | n', t \rangle = \delta_{n,n'}
$$
we find by differentiating with respect to $t$ and using (\ref{diffHop})
that the coefficients $\alpha_1$ and $\beta_1$ must be purely imaginary. 
Finally, we consider a general state vector in the Heisenberg picture,
\begin{equation}   \label{psiH}
   |\psi\rangle = \sum_n \psi(n,t) \, | n , t \rangle
\end{equation}
We take ${\rm d}/{\rm d}t$, use (\ref{diffHop}), put $x=na$, and reexpress 
$\alpha_1$ and $\beta_1$ by 
$$ 
     U=(\alpha_1 + 2\beta_1) \, i\hbar  \qquad \qquad   
        \frac1{2m}=\frac{a^2\beta_1}{i\hbar}
$$  
Thus we find
$$
    i \hbar \frac{\rm d}{{\rm d}t} \psi(x,t) = U \psi(x,t) 
  - \frac{\hbar^2}{2m} \frac{\psi(x+a,t)+\psi(x-a,t)-2\psi(x,t)}{a^2}
$$
This equation converges to the free Schr\"odinger equation in the 
continuum limit $a\to 0$.

In Section 2, the hopping-parameter description of a Schr\"odinger particle
is discussed in full generality. 
Hopping amplitudes will not be restricted to next neighbours, 
and it will only be assumed that the hopping 
amplitudes realise the full translational and cubical symmetries of the lattice in 
${\cal O}(1/a^2)$ while any inhomogeneities in the hopping process are at most of 
${\cal O}(1/a)$. Then a (trivial) renormalization scheme exists for the 
continuum limit $a\rightarrow0$ which leads to the standard nonrelativistic 
Schr\"odinger equation, with a vector potential and a scalar potential.  

In Section 3, the hopping-parameter approach is applied to quantum 
electrodynamics without charges and currents. This requires the discretization 
of both, the values of a field $u(x)$ and its spatial variable $x$. The reader
of section 3 is assumed to be somewhat familiar with lattice gauge theory
\cite{Creutz}.
In fact, the model considered in this section is a Hamiltonian version of the
intensively studied $Z(N)$ lattice gauge theory \cite{CrJaRe}.
The Hamilton operator
of the electromagnetic field is recovered in the twofold limit of $N\to\infty$
and zero lattice spacing. Section 4 contains some concluding remarks.

\section{Schr\"odinger particle in 3 dimensions}

Consider a simple cubic lattice where $\vec{x}=a\vec{n}$ 
is the position vector of a site, $a$ is the lattice spacing, and 
$\vec{n}$ an integer vector. 
The most general hopping equation for a single-component wave function as 
defined in (\ref{psiH}) is
\begin{equation}    \label{SchrDiskrGen}
 i \hbar \frac{{\rm d}}{{\rm d}t} \, \psi(\vec{x},t) 
         =  \sum_{\vec{n}} 
            \kappa(\vec{x},\vec{n},t) \, \psi(\vec{x}+a\vec{n},t) 
\end{equation} 
The factor of $i\hbar$ is only cosmetic, since the hopping parameters 
$\kappa(\vec{x},\vec{n})$ can be any complex numbers, so far. 
Conservation of probability requires
\begin{equation}    \label{kappa-}
   \kappa(\vec{x}-a\vec{n},\vec{n},t) = \overline{\kappa(\vec{x},-\vec{n},t)} 
\end{equation}
An important case of reference is that of a free particle, characterized by
hopping parameters with the full symmetry of the lattice.
Then $\kappa(\vec{x},\vec{n},t) = \kappa_0(\vec{n})$ because of translational 
invariances. Cubic symmetry implies 
\begin{equation}    \label{nkappa}
     \kappa_0(\vec{n}) = \kappa_0(-\vec{n})
\end{equation}
so that all $\kappa_0(\vec{n})$ are real because of (\ref{kappa-}). 
Most importantly, the symmetry also implies 
$ \sum_{\vec{n}} \kappa_0(\vec{n}) \, n_i n_j \propto \delta_{ij}$. 
A convenient parametrization is
\begin{equation}   \label{kappaKub}
  \sum_{\vec{n}} \kappa_0(\vec{n}) \, n_i n_j 
   = - \frac{\hbar^2}{m a^2} ~ \delta_{ij} 
\end{equation}
The reduced parameter $m$ will be identified as the particle mass later on; 
the sign of $m$ is discussed in the Conclusions. 
In general, the sum in equation (\ref{kappaKub}) need not converge. 
Assuming convergence here is the basis for the nonrelativistic physics
as it emerges in the form of the Schr\"odinger equation in the continuum limit.

To recover the Schr\"odinger equation, we Taylor-expand the displaced wave functions
on the {\sc rhs} of (\ref{SchrDiskrGen}),
\begin{equation} \label{TaylorPsi} 
   \psi(\vec{x}+a\vec{n},t) = \psi(\vec{x},t) + a n_i \nabla_i \psi(\vec{x},t)
    + {\textstyle\frac12} a^2 n_i n_j \nabla_i \nabla_j \psi(\vec{x},t) 
    + {\cal O}(a^3)
\end{equation} 
Again, let us consider a free particle first.  
Inserting $\kappa(\vec{x},\vec{n},t) = \kappa_0(\vec{n})$ in (\ref{SchrDiskrGen})
and using (\ref{TaylorPsi}), (\ref{nkappa}), and (\ref{kappaKub}) we find
\begin{equation} \label{freeSchr}
 i \hbar \frac{{\rm d}}{{\rm d}t} \, \psi(\vec{x},t) 
         = E_0 \, \psi(\vec{x},t)
  - \frac{\hbar^2}{2m} \vec{\nabla}\cdot\vec{\nabla} \psi(\vec{x},t)  + {\cal O}(a)
\end{equation}
where $E_0 = \sum_{\vec{n}} \kappa_0(\vec{n})$ is certainly infinite but does not 
affect the shape of the wavefunctions. In contrast, the parameter $m$ determines the 
particle mass and must be finite, as anticipated in definition (\ref{kappaKub}). 
 
Now we ``turn on'' deviations of the hopping parameters from 
$\kappa_0(\vec{n})$. Let us put
\begin{equation}     \label{k_0+k_1}
    \kappa(\vec{x},\vec{n},t) = \kappa_0(\vec{n}) + \kappa_1(\vec{x},\vec{n},t)
\end{equation}
Again, we insert (\ref{TaylorPsi}) in (\ref{SchrDiskrGen}).
The multiplicative terms on the {\sc rhs} of (\ref{SchrDiskrGen}) now are
$ E_0 \psi(\vec{x},t) + 
   \sum_{\vec{n}}  \kappa_1(\vec{x},\vec{n},t) ~ \psi(\vec{x},t) $.
The inhomogeneous term can be rewritten as
$$  
   \frac12 \sum_{\vec{n}} \left( \kappa_1(\vec{x},\vec{n},t) +
      \kappa_1(\vec{x},-\vec{n},t)\right) ~ \psi(\vec{x},t)      
$$
Using (\ref{kappa-}) and expanding the ensuing displaced argument, 
we obtain the following form of the multiplication operator: 
$$
  \frac12 \sum_{\vec{n}} \left( \kappa_1(\vec{x},\vec{n},t) -
 \overline{\kappa_1(\vec{x},\vec{n},t)} \right) + 
  \frac12 \sum_{\vec{n}} a \vec{n}\cdot
      \vec{\nabla}\overline{\kappa_1(\vec{x},\vec{n},t)} 
 + {\cal O}(a^2 \kappa_1)
$$
This shows that for a finite $\vec{x}$-dependent contribution,
the real part of $\kappa_1$ must be of ${\cal O}(1)$ while the imaginary 
part can be of ${\cal O}(1/a)$. Hence, if we define a vector potential
\begin{equation} \label{defA}
     \vec{A}(\vec{x},t) = 
  \frac{ma}{e\hbar} \sum_{\vec{n}} \vec{n} \, \Im\kappa_1(\vec{x},\vec{n},t)  
\end{equation} 
then the multiplicative terms of (\ref{SchrDiskrGen}) take the form 
\begin{equation} \label{locMul}
 \left( E_0 + \sum_{\vec{n}} \Re \kappa_1(\vec{x},\vec{n},t) \right) \psi(\vec{x},t) 
  + i \frac{e\hbar}{2m} \left( \vec{\nabla}\cdot\vec{A}(\vec{x},t) \right) 
                                                                \psi(\vec{x},t)   
\end{equation}
The gradient terms on the {\sc rhs} of (\ref{SchrDiskrGen}) 
can be written as
$$
    \frac{a}2 \vec{\nabla}\psi(\vec{x},t) \cdot \sum_{\vec{n}} \vec{n} 
    \left( \kappa(\vec{x},\vec{n},t) - \kappa(\vec{x},-\vec{n},t) \right) 
$$
By (\ref{k_0+k_1}) and (\ref{kappa-}) this is equal to 
$$
    \frac{a}2 \vec{\nabla}\psi(\vec{x},t) \cdot \sum_{\vec{n}} \vec{n}  
    \left( \kappa_1(\vec{x},\vec{n},t) 
  - \overline{\kappa_1(\vec{x}-a\vec{n},\vec{n},t)} \right)    
$$
The displacement of $\vec{x}$ in 
$\overline{\kappa_1(\vec{x}-a\vec{n},\vec{n},t)}$ 
produces a term of higher order in $a$ which can be neglected in the limit 
$a\to 0$. Thus the only relevant contribution to the 
gradient terms comes from the imaginary part of $\kappa_1(\vec{x},\vec{n},t)$ 
and is of the form
\begin{equation} \label{gradTerm}
  i \frac{e\hbar}{m}
  \left( \vec{\nabla}\psi(\vec{x},t) \right) \cdot \vec{A}(\vec{x},t) 
\end{equation}
where $\vec{A}(\vec{x},t)$ is the same as in (\ref{defA}).

With inhomogeneities of ${\cal O}(1)$ in the real part, and of ${\cal O}(1/a)$
in the imaginary part, it is clear that the double-gradient terms of equation
(\ref{SchrDiskrGen}) are the same as in the free-particle case (\ref{freeSchr}).
Collecting all the terms discussed above, we recover from (\ref{SchrDiskrGen})
the general, nonrelativistic Schr\"odinger equation 
\begin{equation}   \label{genSchr}
    i\hbar\frac{\partial}{\partial t}\psi(\vec{x},t) = 
    \frac{1}{2m}  
  \left(\frac{\hbar}{i}\vec{\nabla} - e\vec{A}(\vec{x},t)\right)^2  
  \psi(\vec{x},t) + U(\vec{x},t) \psi(\vec{x},t)  
\end{equation}
with the vector potential of equation (\ref{defA}) and the scalar potential 
\begin{equation}   \label{defU}
    U(\vec{x},t) = E_0 + \sum_{\vec{n}} \Re \kappa_1(\vec{x},\vec{n},t)
                                     - \frac{e^2}{2m} \vec{A}(\vec{x},t)^2 
\end{equation}
In canonical quantization, the prescription is to identify $U(\vec{x},t)$ and 
$\vec{A}(\vec{x},t)$ with the corresponding functions of the classical Hamiltonian. 
This amounts to an extrapolation into the microscopic domain. The corresponding
procedure in the present context is as follows. By Ehrenfest's theorem, eq.\ 
(\ref{genSchr}) will reproduce the classical equations of motion for the centre of 
a wave packet in the limit $\hbar\to 0$. The classical $U(\vec{x},t)$ and 
$\vec{A}(\vec{x},t)$ then coincide with those in the Schr\"odinger equation. 
Thus, if desired, $U(\vec{x},t)$ and $\vec{A}(\vec{x},t)$ can be extrapolated 
as with canonical quantization. 

In concluding the section, it should be noted that the order-of-magnitude
assumptions for the hopping parameters depend on the further assumption that no
dramatic cancellations occur between $\kappa(\vec{x},\vec{n},t)$ for different
$\vec{n}$. Of course, those cancellations would require some extra reason for a 
fine-tuning.
In the absence of a reason, the assumptions describe the most general and,
hence, the most likely set of parameters consistent with the constraints.

\section{Free electromagnetic field}

This section is to demonstrate that ``unitary hopping'' can be a useful concept 
also for quantum field theories. We here consider source-free $U(1)$ 
gauge theory. Its Hamilton operator in the temporal gauge \cite{Kogut,ChristLee} 
is an $\infty$-dimensional version of (\ref{genSchr}). 
A ``hopping'' scenario requires the configuration space to be
discrete. Thus local gauge invariance will have to be discretized, too. 
In case of $U(1)$ this can be done in a way that preserves an exact local gauge 
group, namely $Z(N)$, whose limit $N\to\infty$ reproduces $U(1)$.  

In lattice gauge theory, a gauge field lives on the links between next-neighbour 
lattice sites. A link can be specified by the site $\vec{s}=(n_x,n_y,n_z)$ from 
which it emanates in a positive direction, and by the corresponding $k=1,2,3$. 
In $Z(N)$ gauge theory \cite{CrJaRe} the link variables are phase factors of the 
form
\begin{equation}  \label{varZ}
   e^{2\pi i l/N} \qquad l = 0,1,\ldots,N-1
\end{equation}
They are related to the electromagnetic vector potential $A(\vec{s},k)$, integrated 
along the link, by 
\begin{equation}  \label{phase/A}
   \exp\left( 2\pi i l / N \right) = \exp\left( i a e A / \hbar \right)        
\end{equation} 
Thus a $Z(N)$ gauge field configuration is determined by the numbers 
\begin{equation}  \label{linkNot}
   l(\vec{s},k) \equiv l(n_x,n_y,n_z,k) 
                                       \qquad n_i=0,\pm1,\pm2,\ldots \quad k=1,2,3
\end{equation}
We shall indicate by omitting the arguments $\vec{s}$ and $k$ that we mean the 
configuration as a whole.

The Hamiltonian will be postulated below to be invariant under charge conjugation 
$\cal C$, and under space inversion $\cal P$ about any point $\vec{s}_0$. As it
follows from the relation (\ref{phase/A}) to the vector potentials 
(see also \cite{Schier}), $\cal C$ and $ {\cal P}_{\vec{s}_0} $ are characterized 
by their action on the link variables, 
\begin{eqnarray}     
   {\cal C} \, l(\vec{s},k) & = & -l(\vec{s},k) \label{C}                     \\
   {\cal P}_{\vec{s}_0} \, l(\vec{s},k) & = & -l(2\vec{s}_0-\vec{s}-\hat{k},k) 
                                                \label{P}
\end{eqnarray} 
We also postulate invariance under local $Z(N)$ gauge transformations.
These are characterized by a number $g(\vec{s})=0,1,\ldots,N-1$ on each lattice site.
The link field configuration transforms according to  
$$
   l'(\vec{s},k) = l(\vec{s},k) + g(\vec{s}+\hat{k}) - g(\vec{s}) 
$$
The elementary gauge-invariant construct on a time slice is the plaquette variable
\begin{equation}  \label{discrRot}
  p(\vec{s},i,k) = l(\vec{s},i) + l(\vec{s}+\hat{i},k) 
                                - l(\vec{s}+\hat{k},i) - l(\vec{s},k)
\end{equation}
Gauge-invariant, too, is any shift of a link variable; in particular,
$$
   l(\vec{s},k) \to l(\vec{s},k) \pm 1 \qquad 
   \mbox{if and only if} \qquad
  l'(\vec{s},k) \to l'(\vec{s},k) \pm 1
$$
The gauge field is quantized by assigning a probability amplitude $\psi(l,t)$
to each link-field configuration $l$. For this ``wavefunction'' the general form 
of a unitary-hopping equation is
\begin{equation}   \label{linkhop}
   i \hbar \frac{{\rm d}}{{\rm d}t}\psi(l,t) = \sum_{\Delta l} 
     \kappa(l,\Delta l) \, \psi(l+\Delta l,t)
\end{equation}
{\em Gauge invariance} of the process requires, in the notation of (\ref{discrRot}),
$$
        \kappa(l,\Delta l) = \kappa(p,\Delta l)
$$
{\em Locality} of link interactions is not as uniquely defined---a fact being
utilized with the ``improved actions'' of numerical lattice gauge theories 
\cite{impAct}. We shall only consider the simplest realization of locality, 
assuming 
\begin{itemize}
\item Link-changing processes are independent on different links. 
\item A plaquette can influence a change on its own links, at most.
\end{itemize}
These assumptions correspond to a pre-relativistic, purely spatial notion of
locality---no reference whatsoever is made to the phenomenon of light.  
By the assumption of independence, a change on $k$ links within the same time 
interval ${\rm d}t$ will come with a factor of $({\rm d}t)^k$ and will contribute 
to the time derivative in equation (\ref{linkhop}) only for $k=1$. Thus the sum 
over all link-changes $\Delta l$ reduces to a sum over one-link changes. 
For further simplification, we only consider a change by one unit, corresponding
to {\em nearest-neighbour} hopping in configuration space. Thus (\ref{linkhop}) 
takes the form
\begin{equation}   \label{localSum}
   i  \hbar \frac{{\rm d}}{{\rm d}t}\psi(l,t) = 
     \sum_{{\rm links}\atop\vec{s},i} \sum_{\pm}
     \kappa_{\pm}(p;\vec{s},i) \, \psi(l \pm u_{\vec{s},i},t)
    \stackrel{\rm def}{=} H \psi(l,t)
\end{equation}
where 
$$
 u_{\vec{s},i} = \left\{\begin{array}{cl} 1 & \mbox{ on link } \vec{s},i \\
                                          0 & \mbox{ elsewhere}\end{array} \right.
$$ 
We intend to Taylor-expand the wavefunction. Instead of the derivative
$\partial/\partial l$ on each link we prefer to use the lattice version of the 
functional derivative $\delta/\delta A$ with respect to the vector potential. 
$l$ and $A$ are related through equation (\ref{phase/A}). 
Hence, $\partial/\partial l$ equals the partial derivative 
$(2\pi\hbar/eNa)(\partial/\partial A)$. 
Now $\partial/\partial A$ can be expressed by the functional derivative 
$\delta/\delta A$ essentially by introducing factors so that in the characteristic 
relation $\partial A(\vec{s},i)/\partial A(\vec{s}',i') = \delta_{\vec{s}\vec{s}'}
\delta_{ii'}$ the $\delta_{\vec{s}\vec{s}'}$ is changed into the lattice delta 
function $a^{-3}\delta_{\vec{s}\vec{s}'}$. Thus, 
$$
   \frac{\partial}{\partial l(\vec{s},i)} = \frac{2\pi\hbar a^2}{e N} \, 
                                            \frac{\delta}{\delta A(\vec{s},i)}
$$ 
Expanding the wavefunction up to order $a^4$ we have
\begin{equation}     \label{psiTay}
   \psi(l\pm u_{\vec{s},i},t) = \psi(l,t) \pm 
 \frac{2\pi\hbar a^2}{e N} \, \frac{\delta\psi(l,t)}{\delta A(\vec{s},i)} +
 \frac{2\pi^2\hbar^2 a^4}{e^2 N^2} \, 
                                 \frac{\delta^2\psi(l,t)}{\delta A(\vec{s},i)^2}
\end{equation}
The first-derivative term is immediately discarded if we postulate that the
Hamiltonian be invariant under {\em space inversion} $\cal P$ (cf.\ (\ref{P})). 
This is because the plaquette variables in the hopping amplitudes 
$\kappa^{\pm}(p;\vec{s},i)$ are invariant under $\cal P$ whereas $l$ and hence 
$\partial/\partial A$ changes sign. 

It remains to discuss the multiplicative terms of (\ref{localSum}). 
To expand the hopping amplitudes in a power series in $a$, we note 
\cite{Creutz,CrJaRe} that the magnetic flux density 
$B_i=\frac12\epsilon_{ijk} F_{jk}$ is related to the plaquette variable by  
$$
   \exp\left(i a^2 e F_{jk}(\vec{s}) / \hbar \right) 
 = \exp\left(2\pi i p(\vec{s},j,k) / N       \right)      
$$
Thus, at a given flux density of ${\cal O}(1)$, the plaquette phase factor
deviates from $1$ only in ${\cal O}(a^2)$, while the plaquette variable $p$ is of 
${\cal O}(a^2 N)$. To be on the safe side, we therefore expand the hopping 
amplitude as a function of $a^2 F_{ij}$ instead of $p$.    
Furthermore, we invoke our locality postulates to restrict plaquettes with
an influence on link $(\vec{s},i)$ to the four cases $p(\vec{s},i,j)$ 
and $p(\vec{s}-\hat{j},i,j)$ with $j\neq i$. Thus, expanding 
$\kappa_{\pm}(p;\vec{s},i)$ to ${\cal O}(a^4)$ we obtain
\begin{equation}     \label{kappaTay}
 \kappa^{(0)}_{\pm}(\vec{s},i) 
   + \frac{e a^2}{\hbar} \sum_{j\neq i} \left( \kappa^{(1)}_{\pm}(\vec{s},i,j) 
     F_{ij}(\vec{s}) + \kappa^{(1)\prime}_{\pm}(\vec{s},i,j) F_{ij}(\vec{s}-\hat{j}) 
                \right) 
  \quad +
\end{equation}
$$ \hspace*{40mm} + \quad \frac{e^2 a^4 }{\hbar^2} 
\sum_{j,j'\neq i} \kappa^{(2)}_{\pm}(\vec{s},i,j,j') 
            F_{ij}(\vec{s}) F_{ij'}(\vec{s})       $$
where in the last term we have discarded any shift of $\vec{s}$ by $\hat{j}$ 
or $\hat{j}'$ as this would lead to an ${\cal O}(a^5)$ contribution. 
 
The $a^2$ terms of expression (\ref{kappaTay}) must vanish if the Hamiltonian 
is to be invariant under {\em charge conjugation}. This is because  
$\cal C$ (cf.\ (\ref{C})) reverses the values of both links and plaquettes, hence 
reverses the sign of the $a^2$ term in (\ref{kappaTay}), while all remaining terms 
of (\ref{kappaTay}) and also of (\ref{psiTay}) are $\cal C$-invariant.

By {\em translation} invariance of the hopping process, all $\kappa$'s must be 
independent of the site vector $\vec{s}$. By invariance under {\em reflections}
about a coordinate plane, $\kappa^{(2)}_{\pm}(\vec{s},i,j,j')$ in the $F^2$ term of
(\ref{kappaTay}) must be proportional to $\delta_{jj'}$. Hence, by 
{\em cubic rotational} invariance, it must be independent of $i$.  
For the same reason, $\kappa^{(0)}_{\pm}(i)$ as the relevant coefficient of 
$\delta^2\psi(l,t)/\delta A(\vec{s},i)^2$ must be independent of $i$.   

Inserting in (\ref{localSum}) the remaining terms of (\ref{psiTay}) and 
(\ref{kappaTay}) we identify the Hamiltonian of free QED as
$$
   H = v + \frac12 \sum_{\vec{s}} a^3 \sum_{i} \left( 
      - \frac{\hbar^2}{\epsilon_0}\frac{\delta^2}{\delta A(\vec{s},i)^2} +
    \frac1{\mu_0} B_i^2(\vec{s}) \right) + {\cal O}(a^5)  
$$
where $v =  \sum_{\vec{s},i} ( \kappa^{(0)}_{+} + \kappa^{(0)}_{-}) $ is the vacuum 
energy and where
\begin{equation}     \label{eps_0mu_0}      
   \frac1{\epsilon_0}  = - \frac{4\pi^2 a}{e^2 N^2} 
                               ( \kappa^{(0)}_{+} + \kappa^{(0)}_{-})  
   \qquad \quad  
   \frac1{\mu_0} = \frac{4 e^2 a}{\hbar^2} (\kappa^{(2)}_{+}(1,1)
                                           +\kappa^{(2)}_{-}(1,1)) 
\end{equation}
In the limit $a\to 0$ we put $\vec{x}=a\vec{s}$ and ${\rm d}^3x=a^3$ 
to obtain the familiar form
\begin{equation}     \label{HQED}
      H = v + \frac{\epsilon_0}2 \int \vec{E}^2(\vec{x}) ~ {\rm d}^3x  
        + \frac1{2\mu_0} \int \vec{B}^2(\vec{x}) ~ {\rm d}^3x 
\end{equation}
where
$$   
   E_i(\vec{s}) = \frac{i\hbar}{\epsilon_0} \, \frac{\delta}{\delta A(\vec{s},i)}
$$

\section{Conclusions}

We have derived the Schr\"odinger equation for a nonrelativistic scalar particle
and for the free electromagnetic field, starting out from the superposition 
principle for state vectors, using the statistical interpretation, and exploiting 
spatial symmetries to a large extent. 
The ambition was to avoid any use of the distinctly non-quantal
concept of trajectories, even in the path-integral sense. 

In the case of a free particle, which has all the exploitable symmetries,
the approach taken here should be compared with the general, group-theoretical 
approach to quantum mechanics as exposed, for example, in \cite{Ulfbeck}.
The main difference is that we found it unnecessary to consider any classical 
space-time symmetries (Galilei or Lorentz transformations).
Rather, the structure of the dynamics follows from 
spatial symmetries together with the {\em absence} of motional information 
from states such as $|\vec{x},t\rangle$. That absence {\em induces} symmetries 
of the time evolution which, however, can be realized only by way of a 
superposition.     

As we have seen, Taylor expansions led to 2nd order derivatives 
and, in the case of QED, to the $B^2$ magnetic energy in the Hamiltonian.
The {\em sign} of the Taylor coefficients, though, must be determined by extra 
arguments.    
For the mass parameter $m$ in equation (\ref{kappaKub}), it is a matter of 
convention whether kinetic energies are always taken as positive or always 
negative, so both signs of $m$ would seem to make physical sense.
A similar remark applies to the case of free QED, except for the {\em relative} 
sign of the parameters $\epsilon_0$ and $\mu_0$ in (\ref{eps_0mu_0}). Here an
additional assumption is required, such as the existence of a ground state, to 
recover the positive phenomenological sign.  

For the definition of the mass in (\ref{kappaKub}) it was essential 
that a free particle find identical hopping conditions on every site of the 
lattice. But this is also what characterizes the lattice as a cartesian 
coordinate system. In case of QED, a cartesian structure is comprised 
in the local $Z(N)$ gauge invariance.   
Thus the unitary-hopping scenario may explain why cartesian coordinates 
play such a preferred role in a wide range of quantum systems \cite{ChristLee}.  

Within the ``physical'' subspace of locally gauge-invariant states, the
Hamiltonian dynamics of the electromagnetic field as described by (\ref{HQED})
is automatically Lorentz invariant. This is quite remarkable since we derived
the dynamics from quantum-mechanical principles in which the roles of space and 
time are initially very different. A similar observation was made by 
Bialynicki-Birula \cite{Bialyn} with respect to the Weyl equation.


\begin{thebibliography}{11}

\bibitem{Heisenb}
    W. Heisenberg, {\em Zeitschrift f\"ur Physik}, 49: 619, 1928.

\bibitem{Hubbard} 
                  D. Bareswyl et al.\ (eds.), {\it The Hubbard Model},
                  Proceedings of the Conference on the Mathematics and Physics 
                  of the Hubbard Model (San Sebastian), Plenum, New York 1995. 
\bibitem{Kogut} 
       J. Kogut, L. Susskind, {\em Physical Review} D, 11: 395, 1975.

\bibitem{Creutz} 
        M. Creutz, {\em Quarks, Gluons and Lattices}, Cambridge 1983. 

\bibitem{Bialyn} 
       I. Bialynicki-Birula, {\em Physical Review} D, 49: 6920, 1994. 

\bibitem{Boghosian} 
          B. Boghosian, W. Taylor, {\em Physical Review} E, 57: 54, 1998;
                    {\em Physica} D, 120: 30, 1998; 
                   {\em International Journal of Modern Physics}, 8: 705, 1997.

\bibitem{CrJaRe} 
   M. Creutz, L. Jacobs, C. Rebbi, {\em Physical Review} D, 20: 1915, 1979.

\bibitem{ChristLee} 
   N. H. Christ, T. D. Lee, {\em Physical Review} D, 22: 939, 1980. 

\bibitem{Schier} 
     K. Ishikawa, G. Schierholz, M. Teper, 
        {\em Zeitschrift f\"ur Physik} C, 19: 327, 1983. 

\bibitem{impAct} 
                 T. DeGrand et al.\ (eds.), {\it Lattice '98}, conference
                 proceedings, {\em Nuclear Physics} B (Proc.\ Suppl.), 
                 73, 1999; Section L. 

\bibitem{Ulfbeck} 
          A. Bohr, O. Ulfbeck, {\em Reviews of Modern Physics}, 67: 1, 1995.

\end{thebibliography}
\end{document}